# Comparison of the Performance of Two Service Disciplines for a Shared Bus Multiprocessor with Private Caches

Angel Vassilev Nikolov and Lerato Lerato
[1]Department of Mathematics and Computer Science, National University of Lesotho
Roma, 180, Lesotho

[2] Department of Mathematics and Computer Science, National University of Lesotho
Roma, 180, Lesotho

**Abstract**
In this paper, we compare two analytical models for evaluation of cache coherence overhead of a shared bus multiprocessor with private caches. The models are based on a closed queuing network with different service disciplines. We find that the priority discipline can be used as a lower-level bound. Some numerical results are shown graphically.
.

## 1. Introduction

Caches have been widely used in multiprocessors to improve systems performance. Caching of shared data, however, introduces the cache coherence problem. Simply coherence can be defined as retrieving always the most recent value for any data. Maintaining this feature solely by the software makes the programmer's task extremely difficult. Modern multiprocessors solve the cache coherence problem in hardware by implementing cache coherence protocols [6]. There are two main classes of hardware protocols, snoopy and directory based protocols. Snoopy protocols use broadcast medium and hence apply to a smaller-scale bus-based multiprocessors. In these broadcast systems each cache "snoops" on the bus and watches for transition that affects it. In this paper we consider this class. Coherence requirements can be met in two ways. Invalidate protocols invalidate other cache copies on a write, so the processor has exclusive access to a data before it writes that data. The alternative Update protocols update all the cached copies of the data when that data is written. Most multiprocessors use Invalidate technique rather than Update technique because update transactions are expensive.

Impact on the performance of the cache coherence protocols can be studied using simulation or analytical models. Simulation is accurate but very time consuming. Analytical models based on queuing theory provide simple but approximate approach for estimating the performance of multiprocessors in the early design cycles. The most commonly used method for this purpose is the Mean Value Analysis (MVA), based on the forced law, i.e. in equilibrium the output rate equals input rate. It offers no possibility to study transient behavior, moreover the assumption of exponential service times is not always adequate [3]. Alternative solution is to describe the system using discrete state continuous time Markov processes. In [4] this approach is applied to a priority discipline where the non-blocking (write-back) requests are served immediately after their arrival, and in [5] a First-Come-First-Served (FCFS) discipline is studied. As shown in [4, 5] this method eliminates the main drawbacks of MVA analysis: inability to deal with transients and the constraints on the service time distributions.

## 2. Description of the models

A multiprocessor consists of several processors connected together to a shared main memory by a common complete transaction bus. Each processor has a private cache. When a processor issues a request to its cache, the cache controller examines the state of the cache and takes suitable action, which may include generating bus transaction to access main memory. Coherence is maintained by having all cache controllers "snoop" on the bus and monitor the transaction. Snoopy cache-coherence protocols fall in two major categories: Invalidate and Update [6]. Invalidating protocols are studied here but the concepts can be applied with some modifications to updating protocols too. Transactions may or may not include the memory block and the shared bus. Typical transaction that does not include memory block is Invalidate Cache Copy which occurs when a processor requests writing in the cache. All other processors simply change the status bit(s) of their on copies to Invalid. If the memory block is uncached or not clean it can be uploaded from the main memory, but in todays multiprocessors it is rather uploaded from another cache designated as Owner (O) (cache-to cache transfer). Memory-to cache transfer occurs when the only clean copy is in the main memory. A cache block is written back (WB) in the main memory (bus is used) when a dirty copy is evicted [6]. Apparently the bus can be considered as the bottleneck of the system.

For the model in [4] these WB requests are immediately served, that is they have priority over all other transaction types, and for the model presented in [5], WB requests and all other requests are treated equally, i.e. the service discipline is on First Come First Served (FCFS) basis.

In terms of the queuing theory processors can be viewed as customers (clients) and the bus can be viewed as a server. The FCFS queue and the priority queue are illustrated in Fig. 1.a and Fig 1.b, respectively.

Each processor alternates execution (think, compute) phases and phases when it waits for a memory request to be served. The execution phase is assumed exponentially distributed with parameter $\lambda$. This assumption is adequate for most applications [3]. Immediately after issuing a coherence request the customer blocks itself. The service time for blocking request has a density function $f_1(x)$.

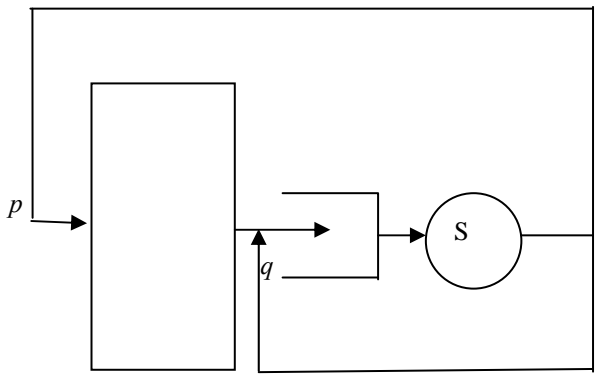

Fig. 1.a

When service is completed the processor (customer) resumes processing with probability $p$ or resumes processing and generates a new request with probability $q$ ($p+q=1$). Details on how to obtain the input parameters are given in [7, 8]. This new request has a different density function $f_2(x)$ and corresponds to WB transaction. It does not block the customer but the server is held until completion of WB transaction therefore adding to the queue. WB request in Fig. 1.a joins the tail of the queue of blocking and non-blocking requests. In Fig. 1.b if a WB request is generated the bus (server) is not relinquished by the processor whose coherence transaction was just completed. The service of the WB request is started immediately for this processor, and only after its completion the first processor in the queue gets access to the bus.

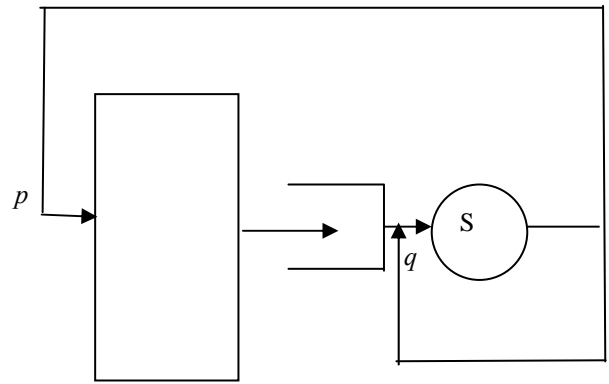

Fig. 1.b

The equations describing these queues and their solutions are given in [4, 5]. We start with fairly complex set of integro-differential equations but the output is a set of linear equations from which the steady-state probabilities and hence the throughput can be determined. For the FCFS discipline, however, the number of linear equations grows enormously for large $N$, so the exact solution is too complicated to be practical. The networks in Fig. 1.a and Fig. 1.b are "work conserving" since the server does not go idle if there is a customer in the queue, and the amount of the service time does not depend on the service discipline [2]. The mean waiting time is same for Fig. 1.a and Fig. 1.b according to the conservation law [4, 5]. Distributions of waits, however, are different: in Fig. 1.b non-blocking (WB) requests do not wait at all because they are served immediately after arrival, so that the waiting time is zero, while for the network in Fig. 1.a the waiting time is greater than zero. Blocking requests in Fig. 1.a therefore wait longer and the throughput is smaller than that in Fig. 1.b, thus we can conclude that the priority scheme (Fig. 1.b) can be regarded as lower-level bound for FCFS discipline (Fig. 1.a).

Average Number of Blocked Customers (processors), *ANBC*, for the two disciplines, computed using the formulas, derived in [4] and [5] are illustrated in Table *1.a* through *1.h* for blocking caches (BL). For all cases $f_1(x) = \mu_1 e^{-\mu_1 x}$ and $f_2(x) = \mu_2 e^{-\mu_2 x}$, and *p=0.8* and $\mu_1 = 0.1$ [1/t.u.][1].

| $\lambda$ | FCFS | Priority | % difference |
|---|---|---|---|
| 0.001 | 0.07014828 | 0.07344077 | 4.69361131 |
| 0.002 | 0.18977920 | 0.20284240 | 6.88336661 |
| 0.003 | 0.34289203 | 0.36860743 | 7.49956231 |
| 0.004 | 0.51524391 | 0.55285868 | 7.30038140 |
| 0.005 | 0.69555280 | 0.74238917 | 6.73368933 |
| 0.006 | 0.87572487 | 0.92861078 | 6.03910016 |
| 0.007 | 1.05045078 | 1.10650487 | 5.33619402 |
| 0.008 | 1.21657142 | 1.27350263 | 4.67964332 |
| 0.009 | 1.37245655 | 1.42859627 | 4.09045568 |
| 0.010 | 1.51749673 | 1.57171551 | 3.57290955 |
| | *1.a)* *N=4, $\mu_2$ =0.01[1/t.u.]* | | |

| $\lambda$ | FCFS | Priority | % difference |
|---|---|---|---|
| 0.001 | 0.10346160 | 0.10967984 | 6.01018450 |
| 0.002 | 0.29814050 | 0.32060606 | 7.53522768 |
| 0.003 | 0.53750740 | 0.57628734 | 7.21477436 |
| 0.004 | 0.78870741 | 0.83859945 | 6.32579880 |
| 0.005 | 1.03240021 | 1.08777002 | 5.36321218 |
| 0.006 | 1.25904105 | 1.31562088 | 4.49388283 |
| 0.007 | 1.46505497 | 1.52011385 | 3.75814410 |
| 0.008 | 1.65010151 | 1.70210131 | 3.15130902 |
| 0.009 | 1.81540668 | 1.86361092 | 2.65528566 |
| 0.010 | 1.96283453 | 2.00700331 | 2.25025446 |
| | *1.b)N=4, $\mu_2$ =0.006667[1/t.u.]* | | |

| $\lambda$ | FCFS | Priority | % difference |
|---|---|---|---|
| 0.001 | 0.09793845 | 0.10393609 | 6.12389310 |
| 0.002 | 0.27607106 | 0.29989579 | 8.62992741 |
| 0.003 | 0.50864514 | 0.55419819 | 8.95576169 |
| 0.004 | 0.77068883 | 0.83443491 | 8.27131302 |
| 0.005 | 1.04195502 | 1.11731297 | 7.23236099 |
| 0.006 | 1.30831803 | 1.38883665 | 6.15436121 |
| 0.007 | 1.56119524 | 1.64188994 | 5.16877695 |
| 0.008 | 1.79617630 | 1.87372614 | 4.31749579 |
| 0.009 | 2.01163620 | 2.08410558 | 3.60250972 |
| 0.010 | 2.20764176 | 2.27409559 | 3.01017303 |
| | *1.c)N=5, $\mu_2$ =0.01[1/t.u.]* | | |

| $\lambda$ | FCFS | Priority | % difference |
|---|---|---|---|
| 0.001 | 0.15052080 | 0.16214798 | 7.72463209 |
| 0.002 | 0.44663225 | 0.48738023 | 9.12338382 |
| 0.003 | 0.80995181 | 0.87613907 | 8.17175362 |
| 0.004 | 1.18338340 | 1.26277196 | 6.70860895 |
| 0.005 | 1.53527328 | 1.61736280 | 5.34689961 |
| 0.006 | 1.85242056 | 1.93081123 | 4.23179647 |
| 0.007 | 2.13201057 | 2.20360391 | 3.35802017 |
| 0.008 | 2.37611381 | 2.43985641 | 2.68264077 |
| 0.009 | 2.58864894 | 2.64458879 | 2.16096717 |
| 0.010 | 2.77390615 | 2.82261385 | 1.75592431 |
| | *1.d) N=5, $\mu_2$ =0.006667[1/t.u]* | | |

| $\lambda$ | FCFS | Priority | % difference |
|---|---|---|---|
| 0.001 | 0.13018580 | 0.13994358 | 7.49526844 |
| 0.002 | 0.38029905 | 0.41883309 | 10.13256380 |
| 0.003 | 0.71209440 | 0.78311783 | 9.97387757 |
| 0.004 | 1.08482965 | 1.17914296 | 8.69383567 |
| 0.005 | 1.46496035 | 1.56995792 | 7.16726391 |
| 0.006 | 1.83015922 | 1.93551924 | 5.75687720 |
| 0.007 | 2.16831417 | 2.26748158 | 4.57347954 |
| 0.008 | 2.47457869 | 2.56422251 | 3.62258949 |
| 0.009 | 2.74850736 | 2.82748357 | 2.87342179 |
| 0.010 | 2.99197323 | 3.06042246 | 2.28776217 |
| | *1.e)N=6, $\mu_2$ =0.01[1/t.u.]* | | |

| $\lambda$ | FCFS | Priority | % difference |
|---|---|---|---|
| 0.001 | 0.20681137 | 0.22608037 | 9.31718650 |
| 0.002 | 0.62839527 | 0.69329368 | 10.32764113 |
| 0.003 | 1.14251792 | 1.24085044 | 8.60665053 |
| 0.004 | 1.65787655 | 1.76701584 | 6.58307686 |
| 0.005 | 2.12793843 | 2.23247579 | 4.91261212 |
| 0.006 | 2.53767126 | 2.63056168 | 3.66045914 |
| 0.007 | 2.88784045 | 2.96720787 | 2.74833134 |
| 0.008 | 3.18525763 | 3.25171044 | 2.08626186 |
| 0.009 | 3.43809333 | 3.49318909 | 1.60250898 |
| 0.010 | 3.65399343 | 3.69949483 | 1.24525116 |
| | *1.f)N=6, $\mu_2$ =0.006666[1/t.u.]* | | |

---

[1] t. u.-time unit

| λ | FCFS | Priority | % difference |
|---|---|---|---|
| 0.001 | 0.16711501 | 0.18183576 | 8.80875460 |
| 0.002 | 0.50408229 | 0.56151276 | 11.39307364 |
| 0.003 | 0.95693912 | 1.05819964 | 10.58170982 |
| 0.004 | 1.46217043 | 1.58867062 | 8.65153531 |
| 0.005 | 1.96736551 | 2.09885671 | 6.68361840 |
| 0.006 | 2.44010356 | 2.56304970 | 5.03856237 |
| 0.007 | 2.86557429 | 2.97351291 | 3.76673603 |
| 0.008 | 3.24038821 | 3.33162878 | 2.81572935 |
| 0.009 | 3.56716080 | 3.64255966 | 2.11369397 |
| 0.01 | 3.85102404 | 3.91251645 | 1.59678065 |
|  | 1.g) $N=7$, $\mu_2=0.01[1/t.u.]$ | | |

| λ | FCFS | Priority | % difference |
|---|---|---|---|
| 0.001 | 0.27284496 | 0.30228227 | 10.78902581 |
| 0.002 | 0.84595961 | 0.94043547 | 11.16789163 |
| 0.003 | 1.53800360 | 1.67035363 | 8.60531312 |
| 0.004 | 2.21148627 | 2.34637907 | 6.09964432 |
| 0.005 | 2.80406596 | 2.92305851 | 4.24357205 |
| 0.006 | 3.30291100 | 3.40088073 | 2.96616340 |
| 0.007 | 3.71635507 | 3.79441291 | 2.10038698 |
| 0.008 | 4.05853034 | 4.11982682 | 1.51031226 |
| 0.009 | 4.34322475 | 4.39112053 | 1.10276991 |
| 0.010 | 4.58205090 | 4.61947983 | 0.81685963 |
|  | 1.h) $N=7$, $\mu_2=0.006667[1/t.u.]$ | | |

Table 1

Since the percentage difference of *ANBC*s is always positive we can confirm that the priority scheme can serve as a lower-level bound.

If we look more closely at the tables, we find that the difference is smaller for heavier workload *(λ)*. In spite of the fact that FCFS is more favorable to shorter request than the priority scheme its impact is diminished if the system handles more requests in the case of heavy workload.

It also can be observed that the difference does not vary significantly with *N*.

## 3. Some numerical results

We measure the system performance in *ANPEC* (Average Number of Processors Engaged in Computation) [1]. Obviously from the definition $ANPEC = N - ANBC$. Results are illustrated in Fig. 2.

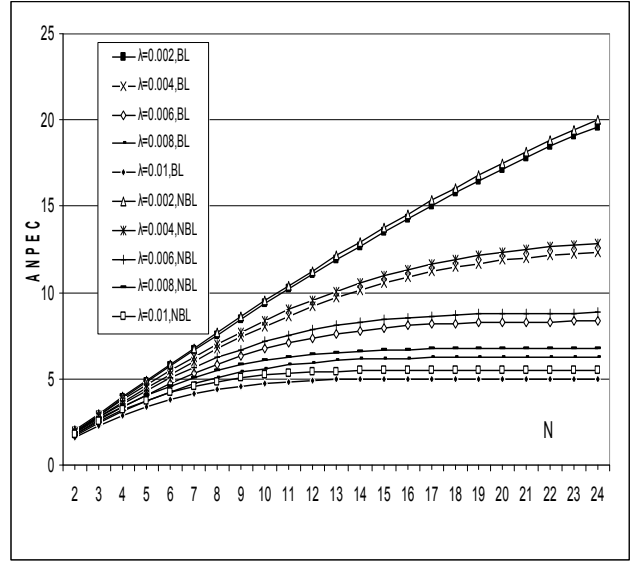

Fig. 2.a. *p=0.9*, $\mu_2=0.01[1/t.u.]$

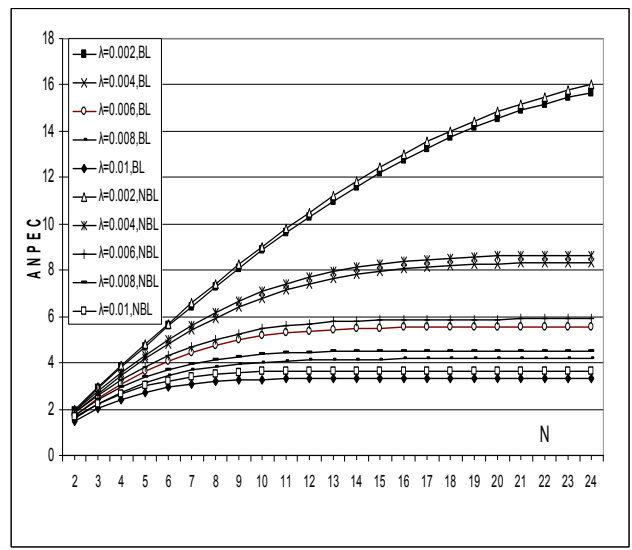

Fig. 2.b. *p=0.8*, $\mu_2=0.01[1/t.u.]$

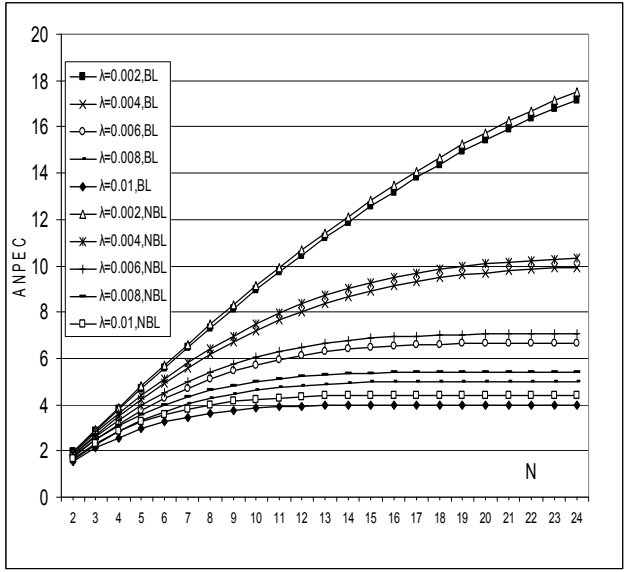

Fig. 2.*c. p=0.9, $\mu_2$ =0.0066666667[1/t.u.]*

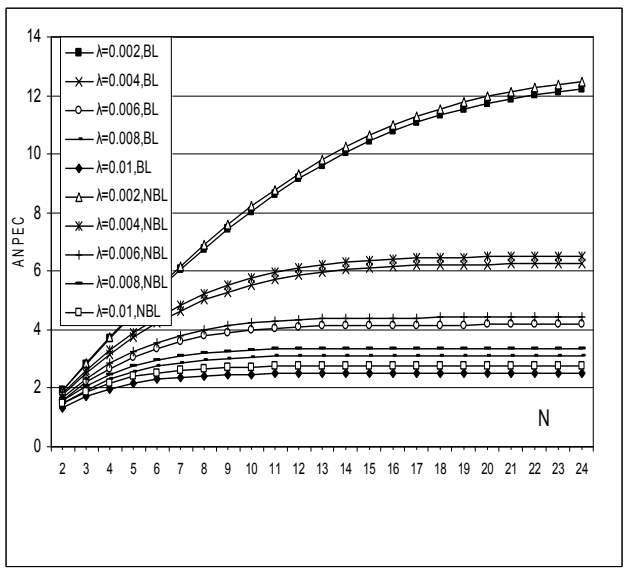

Fig. 2.*d. p=0.8, $\mu_2$ =0.0066666667[1/t.u.]*

From the figures, it can be seen that the performance increases nonlinearly as *N* increases and at some point saturation sets in. Saturation depends heavily on the workload *(λ)*: in all graphics ANPEC saturates quickly with *λ*=0.01[1/t.u.], while it still continues to grow with *λ*=0.001[1/t.u.]. Evidently, with increased main memory traffic (smaller *p*), the performance degradation is more significant (Fig. 2.a and 2.b, and Fig. 2.c and Fig. 2.d). It can also be concluded that the impact of memory access time is also significant, for instance saturation for *λ*=0.01[1/t.u.], and $\mu_2$=0.01[1/t.u.] sets in for *N*=8 (Fig. 2.b), while for $\mu_2$ =0.00666666667[1/t.u.], and same value of *λ* (Fig. 2.d) it occurs for *N*=6.

Apparently, introduction of NBL caches results in improved overall performance.

## 4. Concluding Remarks

Based on the work conservation law we conclude that the priority service discipline produces smaller performance than the FCFS. At the early stage of the design this model can be used as a worst-case approximation for the systems performance. Solving these equations requires insignificant computational effort because their number is *2N+1* [5].

## References

1. A. Ametistova, and I. Mitrani, Modeling and Evaluation of Cache Coherence Protocols in Multiprocessor Systems, *In 9th UK Performance Engineering Workshop for Computer and Telecommunication Systems: Computer and Telecommunication Systems Performance Engineering*, Loughborough University, UK, Jul 1993

2. L. Kleinrock, Queueing Systems, Volume 1, Theory, Wiley-Interscience, 1st Edition, 1975

3. R. E. Matick, Comparison of analytic performance models using closed mean-value analysis versus open-queuing theory for estimating cycles per instruction of memory hierarchies, *IBM Journal of Research and Development*, Jul 2003

4. A. V. Nikolov; Analytical Model for a Multiprocessor with Private Caches and Shared Memory, *Int. Journal of Computers, Communications & Control*, Vol. III (2008), No. 2, pp. 172-182

5. A. V. Nikolov, Model of a Shared-Memory Multiprocessor, *International Journal of Computer Science and Network Security*, vol. 9, No.5, May 2009, pp. 64-70

6. J. Sustersic, A. Hurson, Coherence protocol for bus-based and scalable multiprocessors, Internet and wireless distributed computing environments: a survey, *Advances in Computers*, vol.59, 2003, pp. 211-278

7. S.Srbljic, Z.G. Vranesic, M. Stumm, L. Budin, Models for performance Prediction of Cache Coherence Protocols, *Technical Report CSRI-332,* Jul , 1995, Computer Science Research Institute, University of Toronto

8. S.Srbljic, Z.G. Vranesic, M. Stumm, L. Budin, Analytical prediction of Performance for Cache Coherence protocols, *IEEE*

*Transactions on Computers*, Vol.46, Issue 11 (Nov 1997), pp. 1155-1173

**Angel Vassilev Nikolov** received the BEng degree in Electronic and Computer Engineering from the Technical University of Budapest, Hungary in 1974 and the PhD degree in Computer Science from the Bulgarian Academy of Sciences in 1982 where he worked as a Research Associate. In 1989 he was promoted to Associate Research Professor in Bulgaria. Dr Nikolov also served as a Lecturer of Computer Science at the National University of Science and Technology, Bulawayo, Zimbabwe and at the Grande Prairie Regional College, Alberta, Canada and as an Associate Professor at Sharjah College, United Arab Emirates. Currently he works for the National University of Lesotho, Roma, Lesotho. His research interests include computer architecture, performance evaluation of multiprocessors, and reliability modeling.

**Lerato Lerato** obtained a BSc. degree in Electrical Engineering in 2001 and MSc. Eng (Electrical) in 2004 at the University of Cape Town in South Africa. In 2004 he was employed as a Speech Scientist at Intelleca Voice & Mobile (Pty) Ltd. in Johannesburg. He is currently teaching in the Department of Mathematics and Computer Science at the National University of Lesotho. His current research fields largely include speech technology and IVR platforms and computer architecture.